\documentclass[12pt,tightenlines,eqsecnum,floats,aps,amsmath,amssymb,nofootinbib,prd,showpacs]{revtex4}
\usepackage{setspace}
\usepackage{amsmath,amssymb,amsfonts,amsthm}
\usepackage{graphicx}
\usepackage{enumerate} 
\usepackage{colordvi} 

\def\be{\begin{equation}}
\def\ee{\end{equation}}
\def\ba{\begin{eqnarray}}
\def\ea{\end{eqnarray}}

\def\h{\hat }
\def\b{\bar}

\def\cm{\rm cm}

\def\pphi{p_{(\phi)}}

\def\lp{{\ell}_{\rm Pl}}

\def\rcr{\rho_{\mathrm{crit}}}
\def\rmin{\rho_{\mathrm{min}}}
\def\rmax{\rho_{\mathrm{max}}}
\def\b{$\bullet\,\,\,\, $}
\def\f{\frac}

\def\WDW{WDW\,\,}

\usepackage{enumerate}

\usepackage{colordvi}
\newcounter{mnotecount}[section]

\newcommand{\comment}[1]{}

\begin{document}

\title{Singularity Resolution in Loop Quantum Cosmology:\\
 A Brief Overview}

\author{Abhay Ashtekar}
\affiliation{Institute for Gravitation and the Cosmos, Penn State,
University Park, PA 16802, U.S.A.}

\begin{abstract}

A brief summary of the singularity resolution in loop quantum
cosmology of homogeneous isotropic models is presented.%
\footnote{To appear in the proceedings of the conference NEB-XIII:
\emph{Recent Developments in Gravity}, held at Thessaloniki in June
2008. Dedicated to the memory of Basilis Xanthopoulos, a founder of
this series of conferences.}
The article is addressed to relativists who do not specialize in
quantum gravity. For further details, and answers to more technical
asked questions, the reader is directed to the original papers and
to more comprehensive recent reviews.

\end{abstract}


\maketitle

\section{Issue of the Beginning and the End}
 \label{s1}

Over the history of mankind, cosmological paradigms have evolved
in interesting ways. It is illuminating to begin with a long range
historical perspective by recalling paradigms that seemed obvious
and most natural for centuries only to be superseded by radical
shifts.\smallskip

Treatise on Time, the Beginning and the End date back at least
twenty five centuries. Does the flow of time have an objective,
universal meaning beyond human perception? Or, is it fundamentally
only a convenient, and perhaps merely psychological, notion? Did the
physical universe have a finite beginning or has it been evolving
eternally? Leading thinkers across cultures meditated on these
issues and arrived at definite but strikingly different answers. For
example, in the sixth century BCE, Gautama Buddha taught that `a
period of time' is a purely conventional notion, time and space
exist only in relation to our experience, and the universe is
eternal. In the Christian thought, by contrast, the universe had a
finite beginning and there was debate whether time represents
`movement' of  bodies or if it flows only in the soul. In the fourth
century CE, St. Augustine held that time itself started with the
world.

Founding fathers of modern Science from Galileo to Newton continued
to accept that God created the universe. Nonetheless, their work led
to a radical change of paradigm. Before Newton, boundaries between
the absolute and the relative, the true and the apparent and the
mathematical and the common were blurry. Newton rescued time from
the psychological \emph{and} the material world and made it
objective and absolute. It now ran uniformly from the infinite past
to the infinite future. This paradigm became a dogma over centuries.
Philosophers often used it to argue that the universe itself
\emph{had} to be eternal. For, as Immanuel Kant emphasized,
otherwise one could ask ``what was there before?''

General relativity toppled this Newtonian paradigm in one fell
swoop. Now the gravitational field is encoded in space-time
geometry. Since geometry is a dynamical, physical entity, it is now
perfectly feasible for the universe to have had a finite beginning
---the big-bang--- at which not only matter but \emph{space-time
itself} is born.  If space is compact, matter \emph{as well as
space-time} end in the big-crunch singularity. In this respect,
general relativity took us back to St. Augustine's paradigm but in a
detailed, specific and mathematically precise form.  In semi-popular
articles and radio shows, relativists now like to emphasize that the
question ``what was there before?'' is rendered meaningless because
the notions of `before' requires a pre-existing space-time geometry.
We now have a new paradigm, a new dogma: In the Beginning there was
the Big Bang.

However, the very fusion of gravity with geometry now gives rise
to a new tension. In Newtonian (or Minkowskian) physics, a given
physical field could become singular at a space-time point. This
generally implied that the field could not be unambiguously
evolved to the future of that point. However, this singularity had
no effect on the global arena. Since the space-time geometry is
unaffected by matter, it remains intact. Other fields could be
evolved indefinitely. Trouble was limited to the one field which
became ill behaved. However, because gravity is geometry in
general relativity, when the gravitational field becomes singular,
the continuum tares and the space-time itself ends.  There is no
more an arena for other fields to live in. All of physics, as we
know it, comes to an abrupt halt. Physical observables associated
with both matter and geometry simply diverge signalling a
fundamental flaw in our description of Nature. \emph{This is the
new quandary.}

When faced with deep quandaries, one has to carefully analyze the
reasoning that led to the impasse. Typically  the reasoning is
flawed, possibly for subtle reasons. In the present case the
culprit is the premise that general relativity ---with its
representation of space-time as a smooth continuum--- provides an
accurate description of Nature arbitrarily close to the
singularity. For, general relativity completely ignores quantum
physics and over the last century we have learned that quantum
effects become important in the physics of the small. They should
in fact be dominant in parts of the universe where matter
densities become enormous. Thus \emph{the occurrence of the big
bang and other singularities are predictions of general
relativity, precisely in a regime where it is inapplicable!}
Classical physics of general relativity does come to a  halt at
the big-bang and the big crunch. But this is not an indication of
what really happens because use of general relativity near
singularities is an extrapolation which has no physical
justification whatsoever. We need a theory that incorporates not
only the dynamical nature of geometry but also the ramifications
of quantum physics. We need a quantum theory of gravity, a new
paradigm.

These considerations suggest that \emph{singularities of general
relativity are perhaps the most promising gates to physics beyond
Einstein.} They provide a fertile conceptual and technical ground
in our search of the new paradigm. Consider some of the deepest
conceptual questions we face today: the issue of the Beginning and
the end End, the arrow of time, and the puzzle of black hole
information loss. Their resolutions hinge on the true nature of
singularities. In my view, considerable amount of contemporary
confusion about such questions arises from our explicit or
implicit insistence that singularities of general relativity are
true boundaries of space-time; that we can trust causal structure
all the way to these singularities; that notions such as event
horizons are absolute even though changes in the metric in a
Planck scale neighborhood of the singularity can move event
horizons dramatically or even make them disappear altogether
\cite{ph}.

Over the last 2-3 years several classically singular space-times
have been investigated in detail through the lens of loop quantum
gravity (LQG). This is a non-perturbative approach to the
unification of general relativity and quantum physics in which one
takes Einstein's encoding of gravity into geometry seriously and
elevates it to the quantum level \cite{alrev,crbook,ttbook}. One is
thus led to build quantum gravity using \emph{quantum} Riemannian
geometry \cite{almmt,rs,al5,alvol}. Both geometry and matter are
\emph{dynamical} and described \emph{quantum mechanically} from the
start. In particular, then, there is no background space-time.

The kinematical structure of the theory has been firmly established
for some years now. There are also several interesting and concrete
proposals for dynamics (see, in particular
\cite{alrev,crbook,ttbook,spinfoam-rev}). However, in my view there
is still considerable ambiguity and none of the proposals is fully
satisfactory. Nonetheless, over the last 2-3 years, considerable
progress could be made by restricting oneself to subcases where
detailed and explicit analysis is possible. These `mini' and `midi'
superspaces are well adapted to analyze the deep conceptual tensions
discussed above. For, they encompass the most interesting of
classically singular space-times
---Friedman-Lema\^{\i}tre-Robertson-Walker (FLRW) universes with
the big bang singularity and black holes with the
Schwarzschild-type singularity--- and analyze them in detail using
symmetry reduced versions of loop quantum gravity. In all cases
studied so far, classical singularities are naturally resolved and
\emph{the quantum space-time is vastly larger than what general
relativity had us believe.} As a result, there is a new paradigm
to analyze the old questions.

In my talk, I focused on cosmological singularities. The material I
covered is discussed in greater detail in the original papers
\cite{mb1,abl,aps1,aps2,aps3,apsv,aps4,acs,cs} and in more
comprehensive reviews \cite{mb-rev,aa-badhonef}. Discussion of black
hole singularities and the issue of information loss can be found
either in the original papers \cite{ab1,ab2,bv,atv} or in a
comprehensive, recent review addressed to non-experts
\cite{aa-mink}. Here, I will confine myself to a sketch of the
singularity resolution in loop quantum cosmology (LQC). Section
\ref{s2} will provide a conceptual setting, section \ref{s3} will
summarize the main results and section \ref{s4} will present the
outlook.

\section{Conceptual Setting}
\label{s2}

I will restrict myself to the simplest cosmological models: FLRW
space-times with a massless scalar field. I will consider both the
k=0 model and the k=1 model with cosmological constant and comment
on the status of more general models. The simplest models are
instructive because in classical general relativity all their
solutions have a big-bang (and/or big-crunch) singularity.
Therefore, a quantum resolution of these singularities is
non-trivial. It is not difficult to incorporate additional matter
fields and anisotropies.

\begin{figure}[]
  \begin{center}
$a)$\hspace{8cm}$b)$
    \includegraphics[width=3.2in,angle=0]{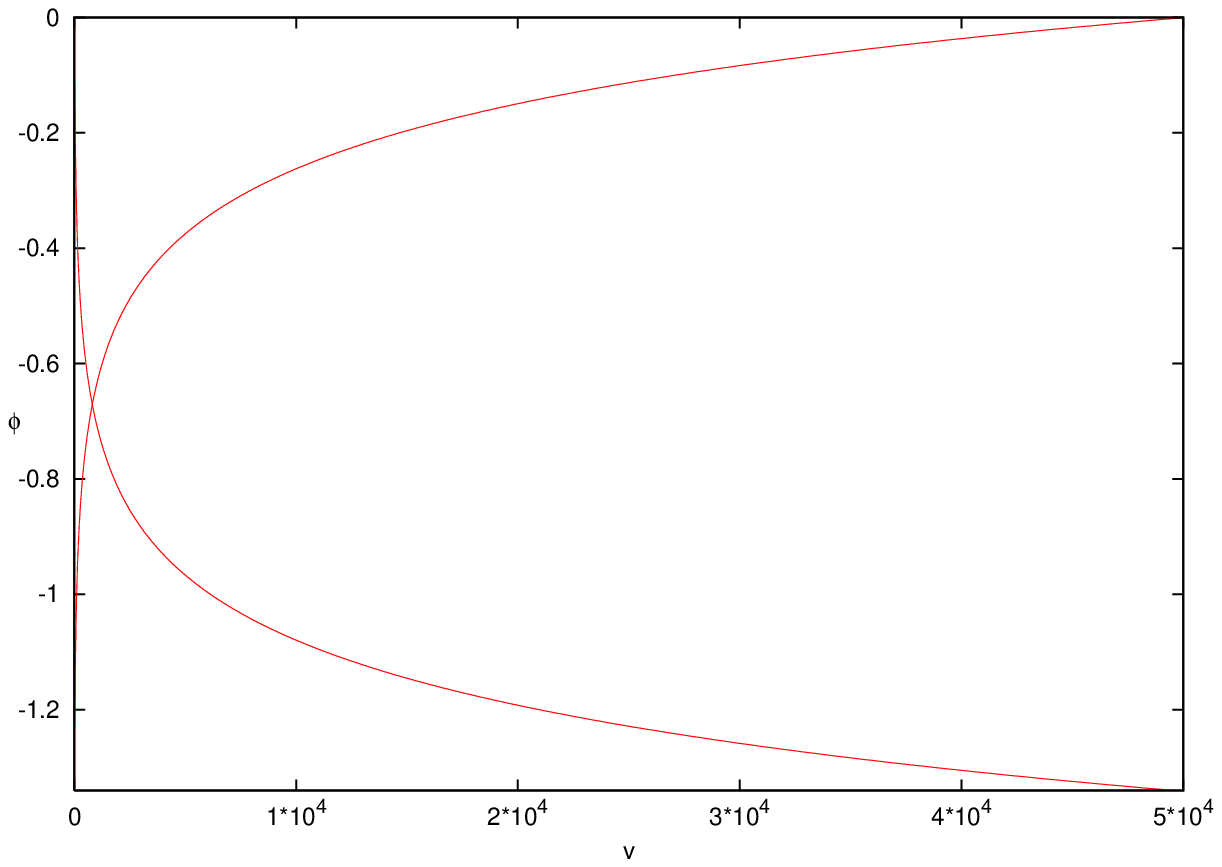}
\includegraphics[width=3.2in,angle=0]{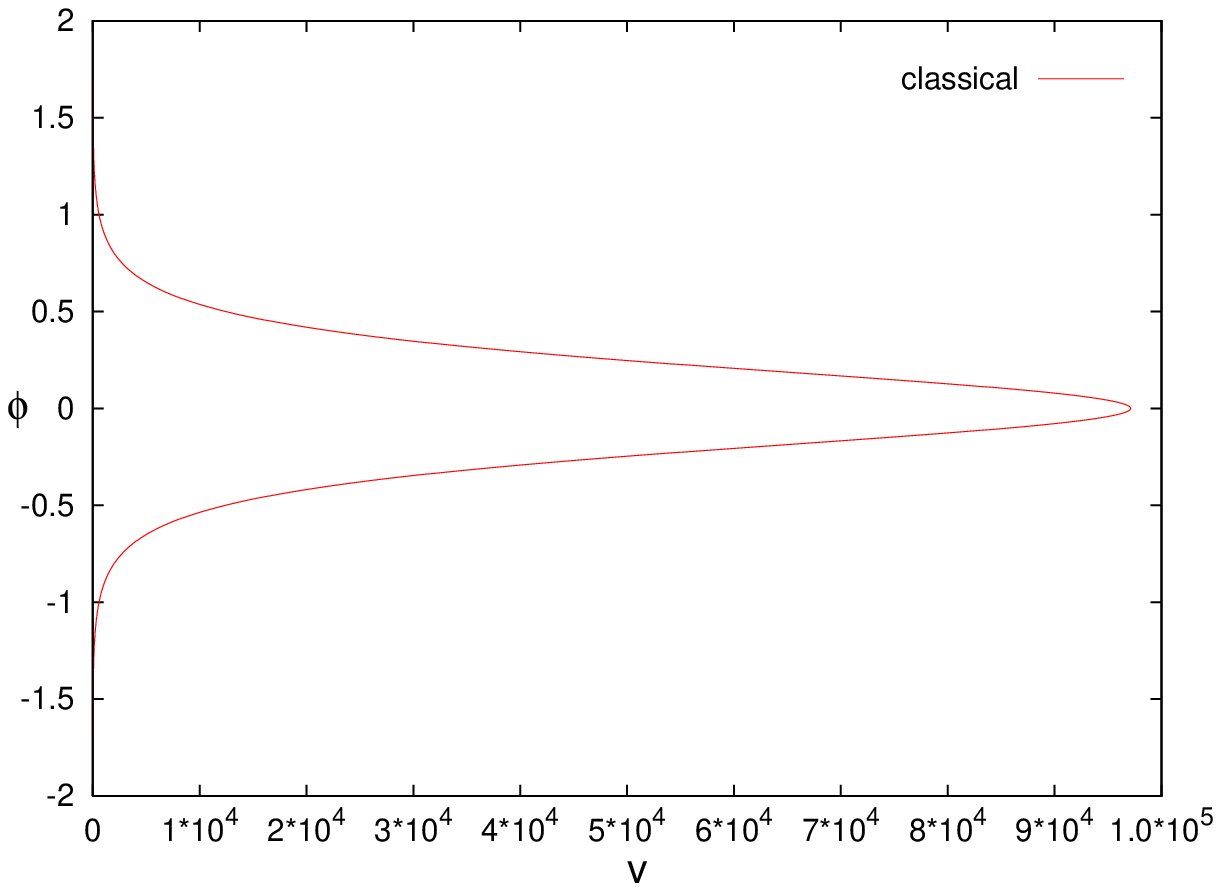}
\caption{Dynamics of FLRW universes with zero cosmological
constant and a massless scalar field. Classical trajectories are
plotted in the $v-\phi$ plane, where $v$ denotes the volume and
$\phi$ the scalar field. $a)$ k=0 trajectories. $b)$ A k=1
trajectory.}\label{class}
  \end{center}
\end{figure}

Figure \ref{class} illustrates classical dynamics for k=0 and k=1
models without a cosmological constant.%
\footnote{In the k=0 case, because the universe is infinite and
homogeneous, to obtain a well-defined Lagrangian or Hamiltonian
formulation, one has to introduce a fiducial box and restrict all
integrations to it. In the k=0 figure, the volume $v$ refers to
this box. Since Lagrangian and Hamiltonian formulations are
stepping stones to the path integral and canonical quantization,
every quantum theory requires a fiducial box. Physical results, of
course, cannot depend on the choice of this box.}
In the k=0 case, there are two classes of trajectories. In one the
universe begins with a big-bang and expands continuously while in
the other it starts out with an infinite volume and contracts
continuously into a big crunch. In the k=1 case, the universe begins
with a big bang, expands to a maximum volume and then undergoes a
recollapse to a big crunch singularity. Now, in quantum gravity, one
does not have a single space-time in the background but rather a
probability amplitude for various space-time geometries. Therefore,
unlike in the classical theory, one cannot readily use, e.g.,  the
proper time along a family of preferred observers as a clock.
However, along each dynamical trajectory, the massless scalar field
$\phi$ is monotonic. Therefore, it serves as a good `clock variable'
with respect to which the physical degrees of freedom (the density
or the volume, anisotropies and other matter fields, if any) evolve.
Note incidentally that, in the k=1 case, volume is double valued and
cannot therefore serve as a global clock variable, while the scalar
field does fulfill this role. The presence of a clock variable is
\emph{not} essential in quantum theory but its availability makes
the relational dynamics easier to grasp in familiar terms. It turns
out that the massless scalar field is well-suited for this purpose
not only in the classical but also in the quantum theory.

If one has a fully developed quantum theory, one can proceed as
follows. Choose a classical dynamical trajectory. Since $\pphi$ is a
constant of motion, this fixes the value of $\pphi$, say $\pphi =
p_{(\phi)}^\star$. Next, choose a point $(v^\star, \phi^\star)$ on
any one of these dynamical trajectories where the matter density and
space-time curvature is low. This point describes the state of the
FLRW universe at a late time where general relativity is expected to
be valid. At the corresponding `internal time' $\phi=\phi^\star$
construct a wave packet which is sharply peaked at $v=v^\star$ and
$\pphi = p_{(\phi)}^\star$ and evolve it backward and forward in
(the scalar field ) time. We are then led to two questions:
\smallskip

i) \emph{The infrared issue:} Does the wave packet remain peaked
on the classical trajectory in the low curvature regime? Or, do
quantum geometry effects accumulate over the cosmological time
scales, causing noticeable deviations from classical general
relativity? In particular, in the k=1 case, is there a recollapse
and if so
for large universes does the value $V_{\rm max}$ of maximum volume
agree with that predicted by general relativity \cite{gu}?

ii) \emph{The ultraviolet issue:} What is the behavior of the
quantum state when the curvature grows and enters the Planck
regime? Is the big-bang singularity resolved? Or, do we need to
supplement dynamics with a new principle, such as the
Hartle-Hawking `no boundary proposal' \cite{hh}? What about the
big-crunch? If they are both resolved, what is on the `other
side'? Does the wave packet simply disperse making space-time
`foamy' or is there a large classical universe also on the other
side? \smallskip

By their very construction, perturbative and effective
descriptions have no problem with the
first requirement. However, physically their implications can not be
trusted at the Planck scale and mathematically they generally fail
to provide a deterministic evolution across the putative
singularity. Since the non-perturbative approaches often start from
deeper ideas, it is conceivable that they could lead to new
structures at the Planck scale which modify the classical dynamics
and resolve the big-bang singularity. But once unleashed, do these
new quantum effects naturally `turn-off' sufficiently fast, away
from the Planck regime? The universe has had some \emph{14 billion
years} to evolve since the putative big bang and even minutest
quantum corrections could accumulate over this huge time period
leading to observable departures from dynamics predicted by general
relativity. Thus, the challenge to quantum gravity theories is to
first create huge quantum effects that are capable of overwhelming
the extreme gravitational attraction produced by matter densities of
some $10^{94}\, {\rm gms/cc}$ near the big bang, and then switching
them off with extreme rapidity as the matter density falls below
this Planck scale. This is a huge burden!

The question then is: How do various approaches fare with respect
to these questions? The older quantum cosmology ---the
Wheeler-DeWitt (WDW) theory--- passes the infra-red test with
flying colors. But unfortunately the state follows the classical
trajectory into the big bang (and in the k=1 case also the big
crunch) singularity. The singularity is not resolved because
\emph{expectation values of density and curvature continue to
diverge} in epochs when their classical counterparts do.

For a number of years, the failure of the \WDW theory to naturally
resolve the big bang singularity was taken to mean that quantum
cosmology cannot, by itself, shed significant light on the quantum
nature of the big bang. Indeed, for systems with a finite number
of degrees of freedom we have the von Neumann uniqueness theorem
which guarantees that quantum kinematics is unique. The only
freedom we have is in factor ordering and this was deemed
insufficient to alter the status-quo provided by the \WDW theory.

The situation changed dramatically in LQG. In contrast to the \WDW
theory, a well established, rigorous kinematical framework \emph{is}
available in full LQG \cite{almmt,alrev,crbook,ttbook}. Furthermore,
this framework is uniquely singled out by the requirement of
diffeomorphism invariance (or background independence)
\cite{lost,cf}. If one mimics it in symmetry reduced models, one
finds that a key assumption of the von-Neumann theorem is violated:
\emph{One is led to new quantum mechanics \cite{abl}!} This quantum
theory is inequivalent to the \WDW theory already at the kinematic
level. Quantum dynamics built in this new arena agrees with the \WDW
theory in `tame' situations but differs dramatically in the Planck
regime, leading to a natural resolution of the big bang singularity.

\section{LQC: Main Results}
\label{s3}

The main LQC results can be summarized as follows
\cite{aps1,aps2,aps3,apsv,aps4,acs,cs}.\smallskip

\begin{figure}[]
  \begin{center}
    $a)$\hspace{8cm}$b)$
    \includegraphics[width=3.2in,angle=0]{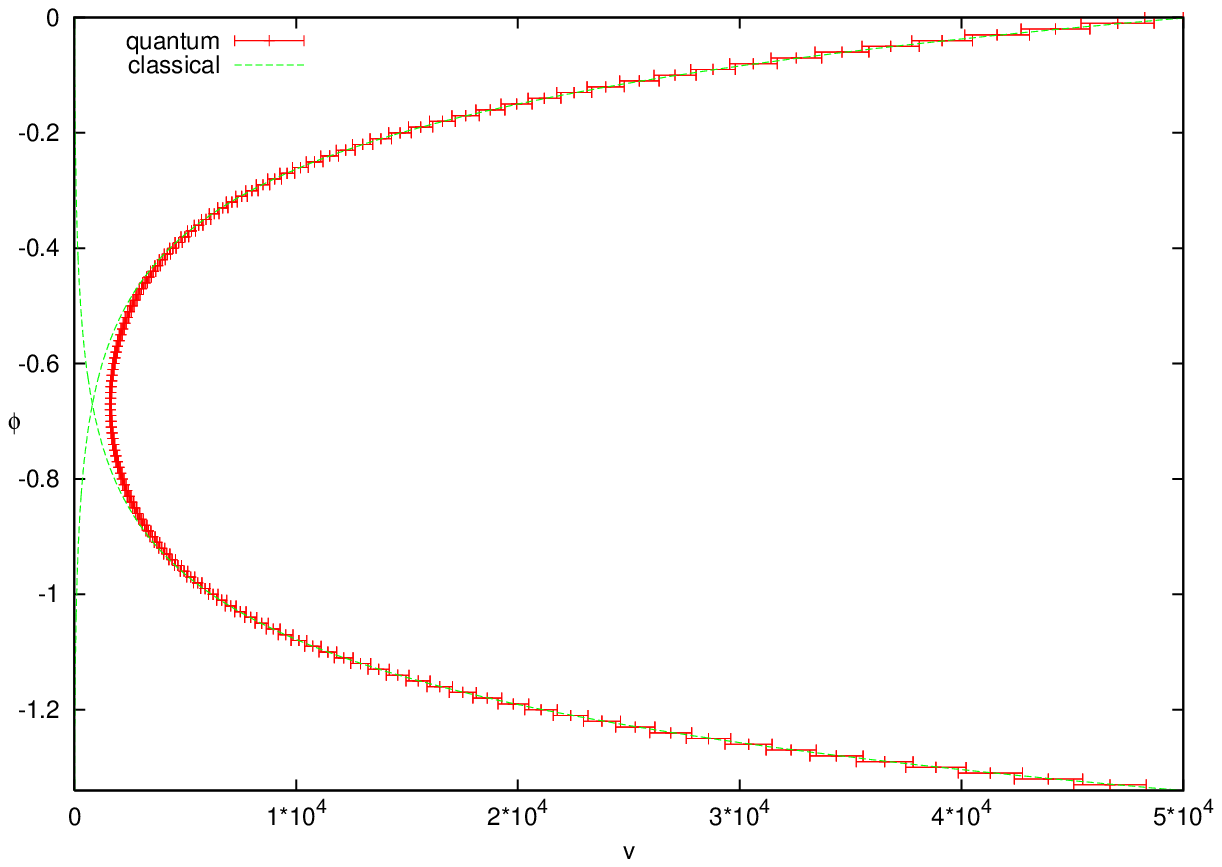}
    \includegraphics[width=3.2in,angle=0]{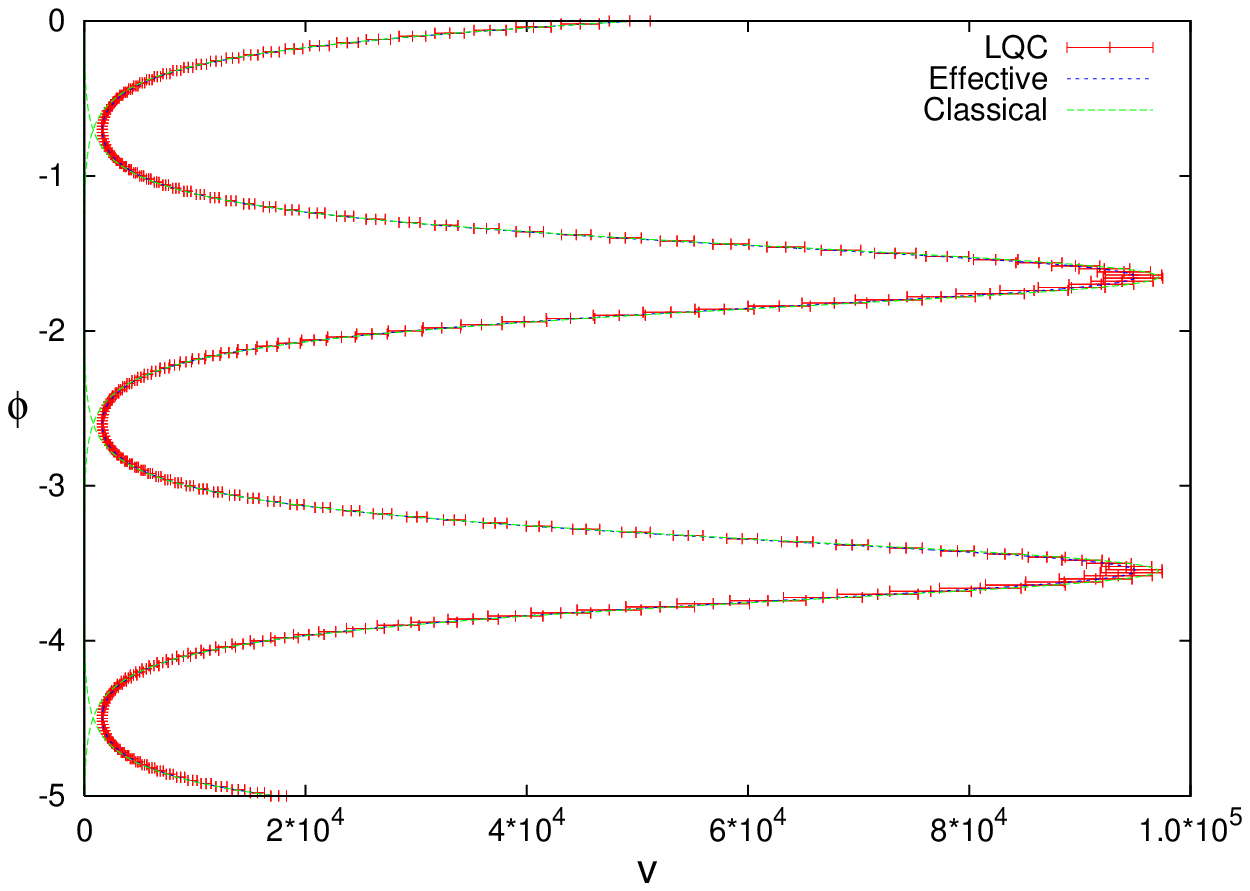}
\caption{In the LQC evolution of models under consideration, the
big bang and big crunch singularities are replaced by quantum
bounces. Expectation values and dispersion of the volume operator
$\h{V}_\phi$, are compared with the classical trajectory and the
trajectory from effective Friedmann dynamics. The classical
trajectory deviates significantly from the quantum evolution at
the Planck scale and evolves into singularities. The effective
trajectory provides an excellent approximation to quantum
evolution at all scales. \,\, $a)$ The k=0 case. In the backward
evolution, the quantum evolution follows our post big-bang branch
at low densities and curvatures but undergoes a quantum bounce at
matter density $\rho \sim 0.41\rho_{\rm PL}$ and joins on to the
classical trajectory that was contracting to the future. $b)$ The
k=1 case. The quantum bounce occurs again at $\rho \sim 0.41
\rho_{\rm Pl}$. Since the big bang and the big crunch
singularities are resolved the evolution undergoes cycles. In this
simulation $p_{(\phi)}^\star = 5\times 10^3$, $\Delta
p_{(\phi)}/p_{(\phi)}^\star = 0.018$, and $v^\star = 5\times
10^4$.} \label{fig:lqc}
\end{center}
\end{figure}

Let us begin with the k=0 model without a cosmological constant.
Following the strategy outlined in section \ref{s2}, let us fix a
point at a late time on
the trajectory corresponding to an expanding classical universe,
construct a
quantum state which is sharply peaked at that point, and evolve it
using the LQC Hamiltonian constraint. One then finds that:\smallskip

\b The wave packet remains sharply peaked on the classical
trajectory till the matter density $\rho$ reaches about 1\% of the
Planck density $\rho_{\rm Pl}$. Thus, as in the \WDW theory, the
LQC evolution meets the infra-red challenge successfully.

\b Let us evolve the quantum state back in time, toward the
singularity. In the classical solution scalar curvature and the
matter energy density keep increasing and eventually diverge at the
big bang. The situation is very different with quantum evolution. As
the density and curvature enter the Planck scale \emph{quantum
geometry effects become dominant creating an effective repulsive
force which rises very quickly, overwhelms the classical
gravitational attraction, and causes a bounce thereby resolving the
big bang singularity.} See Fig \ref{fig:lqc}.

\b At the bounce point, the density acquires its maximum value
$\rmax$. For the class of quantum states under discussion,
numerical simulations have shown that
$\rmax$ is \emph{universal}, $\rho_{\max} \approx 0.41
\rho_{\mathrm{Pl}}$ up to terms $O(\lp^2/a_{\rm min}^2)$,
independently of the details of the state and value of $p_{(\phi)}$
(provided $p_{(\phi)} \gg \hbar$ in the classical units c=G=1).

\b Although in the Planck regime the peak of the wave function
deviates very substantially from the general relativistic
trajectory of figure \ref{class}, it follows an effective
trajectory  with very small fluctuations. This effective
trajectory was derived \cite{jw,vt} using techniques from
geometric quantum mechanics. The effective equation incorporates
the leading corrections from quantum geometry. They modify the
left hand side of Einstein's equations. However, to facilitate
comparison with the standard form of Einstein's equations, one
moves this correction to the right side through an algebraic
manipulation. Then, one finds that the Friedmann equation
$(\dot{a}/a)^2 = (8\pi G\, \rho/3)$ is replaced by
\be \left( \f{\dot{a}}{a} \right)^2 = (8\pi G\,\rho /3)\, \left(1
- \f{\rho}{\rcr}\right) \, . \ee
Here, $\rcr= {\sqrt{3}}/{32\pi^2\gamma^3G^2\hbar}$, where $\gamma$
is the Barbero-Immirzi parameter of LQG (whose value $\gamma \sim
0.24$ is determined by the black hole entropy calculations in
LQG). By plugging in numbers one finds $\rcr \approx 0.41
\rho_{\rm Pl}$. Thus, $\rcr \approx \rmax$, found in numerical
simulations. Furthermore, one can show analytically \cite{acs}
that the spectrum of the density operator on the physical Hilbert
space admits a finite upper bound $\rho_{\rm sup}$, also given by
$\rho_{\rm sup}= {\sqrt{3}}/{32\pi^2\gamma^3G^2\hbar}$. This
result refers to the density operator on the physical Hilbert
space. Thus, \emph{there is an excellent match between quantum
theory which provides $\rho_{\rm sup}$, the effective equations
which provide $\rcr$ and numerical simulations which provide
$\rmax$.}

\b In classical general relativity the right side,\, $8\pi G
\rho/3$,\,
of the Friedmann equation is positive, whence $\dot{a}$ cannot
vanish; the universe either expands forever from the big bang or
contracts into the big crunch. In the LQC effective equation, on the
other hand, $\dot{a}$ vanishes when $\rho=\rcr$ at which a quantum
bounce occurs: To the past of this event, the universe contracts
while to the future, it expands. This is possible because the LQC
correction $\rho/\rcr$ \emph{naturally} comes with a negative sign.
This is non-trivial. In the standard brane world scenario, for
example, Friedmann equation is also receives a $\rho/\rcr$
correction but it comes with a positive sign (unless one
artificially makes the brane tension negative) whence the
singularity is not resolved.

\b Even at the onset of the standard inflationary era, the quantum
correction $\rho/\rcr$ is of the order $10^{-11}$ and hence
completely negligible. Thus, LQC calculations provide \emph{an a
priori justification} for using classical general relativity
during inflation.

\b The analysis has been extended to include a cosmological
constant. In the case when it is negative, the classical universe
starts out with a big bang, expands to a maximum volume and then
undergoes a recollapse to a big crunch singularity. The recollapse
is faithfully reproduced by the LQC evolution. However, in
contrast to general relativity and the \WDW theory, both the
big-bang and the big-crunch singularities are resolved \cite{bp}.
Thus, in this case, the LQC evolution leads to a `cyclic' universe
(as in the k=1 model discussed below).

\b The case of a positive cosmological constant is more subtle
\cite{ap}. Now, as in the $\Lambda=0$ case, the classical theory
admits two types of trajectories. One starts with a big bang and
expands to infinity while the other starts out with infinite
volume and contracts into a big crunch. But, in contrast to the
$\Lambda=0$ case, they attain an infinite volume at a
\emph{finite} value $\phi_{\rm max}$ of $\phi$. The energy density
$\rho|_\phi$ at the `internal time' $\phi$  goes to zero at
$\phi_{\rm max}$. Because the $\phi$ `evolution' is unitary in
LQC, it yields a natural extension of the classical solution
beyond $\phi_{\rm max}$. States which are semi-classical in the
low $\rho|_\phi$ regime again follow an effective trajectory.
Since $\rho|_\phi$ remains bounded, it is convenient to draw these
trajectories in the $\rho_\phi$-$\phi$ plane (rather than
$v$-$\phi$ plane). They agree with the classical trajectories in
the low $\rho_\phi$ regime and analytically continue the classical
trajectories beyond $\rho_{\phi} =0$.

\b The LQC framework has also been extended to incorporate
standard inflationary potential with phenomenologically viable
parameters \cite{aps4}. Again, the singularity is resolved. Thus,
in all these cases, the principal features of the LQC evolution
are robust, including the value of $\rcr$.\\

In the closed, k=1 model, the situation is similar but there are two
additional noteworthy features which reveal surprising properties of
the domain of applicability of classical general relativity.

\b To start with, classical general relativity is again an excellent
approximation to the LQC evolution till matter density $\rho$
becomes about 1\% of the Planck density $\rho_{\rm Pl}$ and, as the
density increases further, the LQC evolution starts departing
significantly. Again, quantum geometry effects become dominant
creating an effective repulsive force which rises very quickly,
overwhelms the classical gravitational attraction, and causes a
bounce thereby resolving both the big bang and the big crunch
singularities. Surprisingly these considerations apply \emph{even to
universes whose maximum radius $a_{\rm max}$ is only} $23 \lp$. For
these universes, general relativity is a very good approximation in
the range $8 \lp < a < 23\lp$. The matter density acquires its
minimum value $\rmin$ at the recollapse. The classical prediction
$\rmin = 3/8\pi Ga^2_{\rm max}$ is correct to one part in $10^{5}$.

\b The volume of the universe acquires its minimum value $V_{\rm
min}$ at the quantum bounce. $V_{\rm min}$ scales linearly with
$p_{(\phi)}$. Consequently, $V_{\rm min}$ can be \emph{much}
larger than the Planck size.  Consider for example a quantum state
describing a universe which attains a maximum radius of a
megaparsec. Then the quantum bounce occurs when the volume reaches
the value $V_{\rm min} \approx 5.7 \times 10^{16}\, {\cm}^3$,
\emph{some $10^{115}$ times the Planck volume.} Deviations from
the classical behavior are triggered when the density or curvature
reaches the Planck scale. The volume can be very large and is not
the relevant scale for quantum gravity effects.

This, in all these cases, classical singularities are replaced by
quantum bounces and LQC provides a rather detailed picture of the
physics in the Planck regime. Furthermore, the singularity
resolution does not cause infra-red problems: There is close
agreement with classical general relativity away from the Planck
scale. The ultraviolet-infrared tension is avoided because,
although quantum geometry effects are truly enormous in the Planck
regime, they die astonishingly quickly.

\section{Outlook}
\label{s4}

Let us summarize the overall situation. In simple cosmological
models, many of the outstanding questions have been answered in
LQC in remarkable detail. The scalar field plays the role of an
internal or emergent time and enables us to interpret the
Hamiltonian constraint as an evolution equation. Singularity is
resolved in a precise sense: While physical observables such as
matter density diverge in classical solutions, giving rise to a
singularity, \emph{they are represented by bounded operators on
the physical Hilbert space}. The big bang and the big crunch are
naturally replaced by quantum bounces. On the `other side' of the
bounce there is again a large universe. General relativity is an
excellent approximation to quantum dynamics once the matter
density falls below one percent of the Planck density. Thus, LQC
successfully meets both the `ultra-violet' and `infra-red'
challenges. Furthermore results obtained in a number of models
using distinct methods re-enforce one another. One is therefore
led to take at least the qualitative findings seriously: \emph{Big
bang is not the Beginning nor the big crunch the End.} Quantum
space-times are vastly larger than what general relativity had us
believe!

How can the quantum space-times of LQC manage to be significantly
larger than those in general relativity when those in the \WDW
theory are not? Main departures from the \WDW theory occur due to
\emph{quantum geometry effects} of LQG. There is no fine tuning of
initial conditions, nor a boundary condition at the singularity,
postulated from outside. Furthermore, matter can satisfy all the
standard energy conditions. Why then does the LQC singularity
resolution not contradict the standard singularity theorems of
Penrose, Hawking and others? These theorems are inapplicable
because \emph{the left hand side} of the classical Einstein's
equations is modified by the quantum geometry corrections of LQC.
What about the more recent singularity theorems that Borde, Guth
and Vilenkin \cite{bgv} proved in the context of inflation? They
do not refer to Einstein's equations. But, motivated by the
eternal inflationary scenario, they assume that the expansion is
positive along any past geodesic. Because of the pre-big-bang
contracting phase, this assumption is violated in the LQC
effective theory.

While the detailed results presented in section \ref{s3} are valid
only for these simplest models, partial results have been obtained
also in more complicated models indicating that the singularity
resolution may be robust \cite{awe2,hybrid,ps}. In this respect
there is a curious similarity with the very discovery of
singularities in general relativity. They were first encountered
in special examples. Although the examples were the physically
most interesting ones ---e.g., the big-bang and the Schwarzschild
curvature singularities--- at first it was thought that these
space-times are singular because they are highly symmetric. It was
believed that generic solutions of Einstein'e equations should be
non-singular. As is well-known, this belief was shattered by the
singularity theorems. Some 40 years later we have come to see that
the big bang and the big crunch singularities are in fact resolved
by quantum geometry effects. Is this an artifact of high symmetry?
Or, are there robust \emph{singularity resolution theorems}
lurking just around the corner?

A qualitative picture that emerges is that the non-perturbative
quantum geometry corrections are \emph{`repulsive'}. While they
are negligible under normal conditions, they dominate when
curvature approaches the Planck scale and can halt the collapse
that would classically have led to a singularity. In this respect,
there is a curious similarity with the situation in the stellar
collapse where a new repulsive force comes into play when the core
approaches a critical density, halting further collapse and
leading to stable white dwarfs and neutron stars. This force, with
its origin in the Fermi-Dirac statistics, is \emph{associated with
the quantum nature of matter}. However, if the total mass of the
star is larger than, say, $5$ solar masses, classical gravity
overwhelms this force. The suggestion from LQC is that a new
repulsive force \emph{associated with the quantum nature of
geometry} comes into play and is strong enough to counter the
classical, gravitational attraction, irrespective of how large the
mass is. It is this force that prevents the formation of
singularities. Since it is negligible until one enters the Planck
regime, predictions of classical relativity on the formation of
trapped surfaces, dynamical and isolated horizons would still
hold. But one cannot conclude that there must be a singularity
because the assumptions of the standard singularity theorems would
be violated. There may be no singularities, no abrupt end to
space-time where physics stops. Non-perturbative, background
independent quantum physics would continue.

At first one might think that, since quantum gravity effects concern
only a tiny region, whatever they may be, their influence on the
global properties of space-time should be negligible whence they
would have almost no bearing on the issue of the Beginning and the
End. However, as we saw, once the singularity is resolved, vast new
regions appear on the `other side' ushering-in new possibilities
that were totally unforeseen in the realm of Minkowski and Einstein.
Which of them are realized generically? Is there a manageable
classification? In the case of black holes, the singularity is again
resolved but there are domains in which geometry is truly quantum:
the quantum fluctuations of the metric operator are so huge near the
putative singularity that the classical field obtained by taking its
expectation value is a poor representation of the actual physics in
these regions \cite{atv}. Presence of such regions would render
classical notions of causality inadequate to understand the global
structure of space-time. In particular, while one can still speak of
marginally trapped surfaces and dynamical and isolated horizons in
the `tame' regions of the full quantum space-time, the notion of an
event horizon turns out to be `too global' to be  meaningful. Is
there perhaps a well-defined but genuinely quantum notion of
causality which reduces to the familiar one on quantum states which
are sharply peaked on a classical geometry? Or, do we just abandon
the idea that space-time geometry dictates causality and formulate
physics primarily in relational terms? There is a plethora of such
exciting challenges. Their scope is vast, they force us to introduce
novel concepts and they lead us to unforeseen territories. These are
just the type of omens that foretell the arrival of a major paradigm
shift to take us beyond
Einstein's space-time continuum.

\bigskip

\textbf{Acknowledgments:}  Much of this chapter is based on the
work with or by Alex Corichi, Tomasz Pawlowski, Param Singh,
Victor Taveras, and Kevin Vandersloot. I have also benefited from
comments, suggestions and probing questions of many colleagues,
especially Martin Bojowald, James Hartle, Jerzy Lewandowski,
Donald Marolf, Roger Penrose, Carlo Rovelli, Bill Unruh and
Madhavan Varadarajan. This work was supported in part by the NSF
grant PHY0456913, the Alexander von Humboldt Foundation, the The
George A. and Margaret M. Downsbrough Endowment and the Eberly
research funds of Penn State.

\end{document}